\documentclass[pra,twocolumn,superscriptaddress,floatfix,amsmath,nofootinbib,amssymb]{revtex4-1}
\usepackage[UKenglish]{babel}
\usepackage{graphicx}
\usepackage{dcolumn}
\usepackage{bm}
\usepackage{verbatim}
\usepackage{mathrsfs}
\usepackage{color}
\newcommand{\ket}[1]{\left| {#1} \right\rangle}

\newcommand{\proj}[2]{\left| {#1} \right\rangle\!\left\langle {#2} \right|}

\newcommand{\ro}{r_{\omega_i}}
\newcommand{\eq}[1]{(\ref{#1})}

\newcommand{\qr}{{q_\text{R}}}
\newcommand{\ql}{{q_\text{L}}}


\begin{document}

\title{Fermionic entanglement extinction in non-inertial frames}
\author{Miguel Montero}
\affiliation{Instituto de F\'{i}sica Fundamental, CSIC, Serrano 113-B, 28006 Madrid, Spain}
\author{Juan Le\'{o}n}
\affiliation{Instituto de F\'{i}sica Fundamental, CSIC, Serrano 113-B, 28006 Madrid, Spain}
\author{Eduardo Mart\'{i}n-Mart\'{i}nez}
\affiliation{Instituto de F\'{i}sica Fundamental, CSIC, Serrano 113-B, 28006 Madrid, Spain}
\begin{abstract}
We study families of fermionic field states in non-inertial frames which show no entanglement survival in the infinite acceleration limit. We generalise some recent results where some particular examples of such states where found. We analyse the abundance and characteristics of the states showing this behaviour and discuss its relation with the statistics of the field. We also consider the phenomenon beyond the single mode approximation.
 
\end{abstract}

\maketitle

\section{Introduction}
One of the aims of the novel field of relativistic quantum information is to study the correlations present between field modes as seen by both inertial and accelerated observers. In this context, there have appeared a number of works studying fermionic fields in non-inertial frames \cite{AlsingSchul,Edu2,Edu3,Edu4,chapucilla,chapucilla2,Shahpoor,Geneferm,chor1,Edu9,chor2}. One of the most well-known scenarios consists in the analysis of the quantum correlations shared between an inertial mode of a fermionic field and an accelerated one, as a function of the  latter's acceleration. Most of the previous literature was centred only in the analysis of pure states from the inertial perspective, where some entanglement was always found to survive in the infinite acceleration limit (see previous citations). This contrasts with the case of bosonic fields, where no field entanglement has ever been found to survive at infinite acceleration limit, and where it is possible even to completely cancel entanglement at any finite acceleration \cite{Alicefalls,Adeschul,Edu9}.

This survival of fermionic entanglement at finite acceleration has been usually linked to the Pauli exclusion principle, which would arguably maintain some entanglement of statistical nature at any value of the acceleration \cite{Edu4,Edu5}. This has led to some works showing that, for some field states, the entanglement at infinite acceleration is not usable for quantum information \cite{Edumucho}.

Nevertheless, in some recent works, fermionic entanglement behaviour for a family of Werner states has been studied \cite{Chinooos} finding fermionic entanglement extinction at finite acceleration for some states of this 1-parametric family. On the other hand, some other works  \cite{iranies} found that if we consider a fermionic tripartite field W-state shared by two accelerated observers, $\text{R}_1$ and $\text{R}_2$  and one inertial, A, i.e.
\begin{align}\ket{\text{W}}=\frac{1}{\sqrt{3}}\left(\ket{100}+\ket{010}+\ket{001}\right),\end{align}
and then for whatever reason we trace out the inertial observer (which is equivalent to considering a system composed of two accelerated partners), then the resulting density matrix
\begin{align}\rho&=\frac{1}{3}\left(\proj{10}{10}+\proj{10}{01}+\proj{01}{10}+\proj{01}{01}\right.\nonumber\\&\left.+\proj{00}{00}\right)\label{ira}\end{align}
is entangled at zero acceleration, but shows no entanglement survival for sufficiently high accelerations of the observers. These particular cases show that the phenomenon of fermionic entanglement survival does not happen for any entangled state. Summarising, working with fermionic fields is a necessary, but not sufficient, condition to observe entanglement survival.

In this work, we study a setting composed of two accelerated observers who watch a general mixed state of two modes of a fermionic field. We generalise previous results about vanishing fermionic entanglement  \cite{iranies,Chinooos} beyond the single mode approximation \cite{Edu9} and  their proper setting, considering both one and two accelereated observers. We also characterise how frequently this phenomenon of entanglement extinction arises by probing the whole space of density matrices rather than restricting to particular families of entangled states. We find that the abundance of mixed states that present the phenomenon is not negligible at all but, instead, it can be about 50\%. Finally we provide a measure of the degree of purity for which the phenomenon starts to manifest.  Our work is structured as follows: In section \ref{set}, we present the setting under consideration and the necessary expressions for the study of field entanglement. Section \ref{res} contains our results about the distribution and characteristics of the phenomenon of vanishing entanglement at the infinite acceleration in general mixed states. Finally, section \ref{cojoniak} contains our conclusions.

\section{Setting}\label{set}
\begin{figure}[hbtp] 
\includegraphics[width=.46\textwidth]{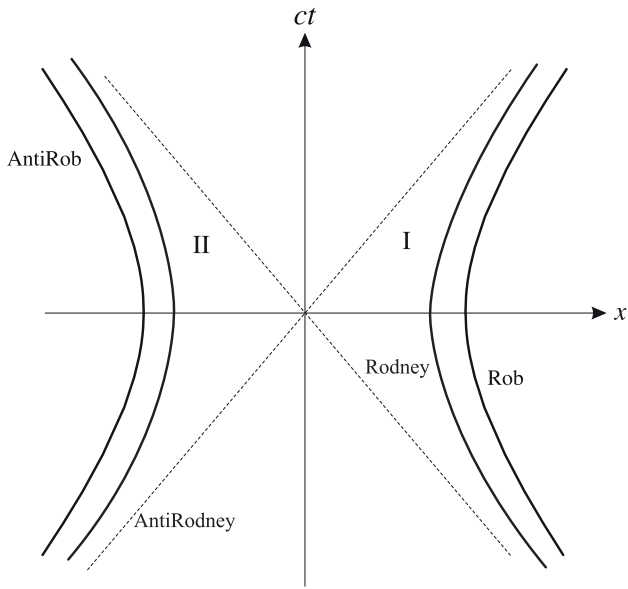}
\caption{Minkowski spacetime diagram showing the world lines of the noninertial observers under consideration. We always consider only two of them at the same time, and label any observer residing in region II with the preffix `Anti'. }
\label{Rindler}
\end{figure}

We consider two non-inertial observers in two causally disconnected patches of Minkowski spacetime. These observers move with uniform proper accelerations, which are are not necessarily equal.  In other words, they travel along lines of constant spacelike Rindler coordinate. We will call the observers Rob or Rodney if they are in region I, and AntiRob or AntiRodney if they are in region II, as depicted in Fig.\ref{Rindler}. Note that although we will refer to the observers by different names depending on the spacetime wedge on which they are, we will always consider only two observers at a time. Each observer probes a single mode of a Grassmann scalar field \cite{AlsingSchul,Edu9}, an anticommuting field with only one degree of freedom. We will analyse entanglement between these modes for different  field states associated with different observers' accelerations.

The most natural way to describe the quantum field from the viewpoint of these uniformly accelerated observers is through field quantisation in the Rindler basis. This means that, whereas an inertial observer would quantise the field using Minkowski modes (positive frequency solutions to the relevant field equation in Minkowski coordinates) to build up the Fock space, it is more appropriate for our observers to quantise the field using Rindler modes (positive frequency solutions in Rindler coordinates). This is so because any particle detector carried by the accelerated observers would couple directly to Rindler rather than Minkowski modes.

The existence of two causally disconnected regions as depicted in Fig. \ref{Rindler} implies that, for each Rindler frequency, there are two different Rindler modes, each of them having support only in either region I or II. When considering a general field state both kinds of modes have to be taken into account.

To study entanglement in non-inertial frames it is convenient to use of the so-called Unruh modes \cite{Edu9}. The annihilation operators associated with these modes have simple expressions in terms of Rindler creation and annihilation operators for particles ($c^\dagger_{\omega,\text{I}},c^\dagger_{\omega,\text{I}}$), and antiparticles ($d^\dagger_{\omega,\text{II}},d^\dagger_{\omega,\text{II}}$), namely
\begin{align}C^\dagger_{\omega,\text{R}}&=\cos r_\omega c^\dagger_{\omega,\text{I}}- \sin r_\omega d_{\omega,\text{II}},\nonumber\\
C^\dagger_{\omega,\text{L}}&=\cos r_\omega c^\dagger_{\omega,\text{II}}- \sin r_\omega d_{\omega,\text{I}},\label{umodes}\end{align}
where $\tan r_\omega=e^{-\pi\omega c/a}$, and $\omega$ indicates Rindler frequency as measured by a Rindler observer with proper acceleration $a$. This acceleration is merely a convention which amounts to the choice of a specific Rindler observer in order to label frequencies; alternatively, we could index the modes by their dimensionless Rindler frequency $\Omega=\omega/a$. The subscripts L and R stand for `left' and `right' modes, which are related to each other by a reversal of regions I and II. These modes have the particularity of being linear combinations of purely positive-frequency Minkowski modes. This implies that the Minkowski and Unruh vacua are the same. Therefore, the Hilbert space factorises as 
\begin{align}\ket{0}_\text{M}=\bigotimes_{\omega}\ket{0}_{\omega,\text{U}}\label{product}\end{align}
where each $\ket{0}_{\omega,\text{U}}$ is annihilated by $C_{\omega,\text{R}}$ and $C_{\omega,\text{L}}$. 

Given \eq{product}, we only need to study detailedly two of the factors in the tensor product, one for each observer, and hence from now on we will consider only the frequencies $\omega_1$ for Rob and $\omega_2$ for Rodney. The relevant part of the vacuum will be written as
\begin{align}\ket{0}_{\text{U}}=\ket{0}_{\omega_1,\text{U}}\ket{0}_{\omega_2,\text{U}}.\label{reduc}\end{align}
We will consider arbitrary Unruh excitations of the form \cite{Edu9}
\begin{align}\ket{1}_{\omega_i,\text{U}}&=C^\dagger_{\omega_i,\text{U}}\ket{0}_\text{U}=\left(\qr_i C^\dagger_{\omega_i,\text{R}}+\ql_i C^\dagger_{\omega_i,\text{L}}\right)\ket{0}_\text{U},\nonumber\\ &\quad \vert\qr_i\vert^2+\vert\ql_i\vert^2=1,\quad i=1,2\label{excs}\end{align}
The case $\qr_i=1$ corresponds to the choice previously known as the single mode approximation \cite{Alsingtelep,AlsingMcmhMil}.

Notice that we have endowed the fermionic Fock space with a particular tensor product structure, one must be careful since the choice of a particular tensor product structure may affect entanglement. We will assume the most extended way of building the fermionic Fock \cite{Edu9} space as it allows us to easily find the relevant expressions for the field states and the main results about entanglement vanishment remain true. For a detailed study of these issues, see \cite{Mig2}.

To study field entanglement, we need to express the Minkowski vacuum and the particle excitations in the Rindler basis. This is easily done from eqs. \eq{umodes} and \eq{reduc} once we impose that the vacuum be annihilated by all four relevant Unruh operators $C_{\omega_1,\text{R}}$, $C_{\omega_1,\text{L}}$, $C_{\omega_2,\text{R}}$, $C_{\omega_2,\text{L}}$. If, following previous notation \cite{Edu9}, we choose a basis for each factor space in \eq{reduc} by
\begin{align}\ket{ijkl}=(c^\dagger_{\omega,\text{I}})^i (d^\dagger_{\omega,\text{II}})^j(d^\dagger_{\omega,\text{I}})^k (c^\dagger_{\omega,\text{II}})^l\ket{0}_\text{Rindler},\end{align}
where $i,j,k,l\in\{0,1\}$ due to Pauli exclusion principle, the Unruh vacuum may be expressed in the Rindler basis through \eq{reduc} with (see \cite{Edu9})
\begin{align}\label{gvac}\ket{0}_{\omega_i,\text{U}}&=\cos^2\ro\ket{0000}-\sin\ro\cos\ro\ket{0011}\nonumber\\&+\sin\ro\cos\ro\ket{1100}-\sin^2\ro\ket{1111}.\end{align}
The excitations can now be straightforwardly computed using \eq{reduc}, \eq{excs} and \eq{gvac}. 

Now we have the states necessary to study field entanglement as seen by two accelerated observers.

One last point remains before the discussion of the formalism is complete: As it is obvious from Fig.\ref{Rindler}, an uniformly accelerated observer lying in one wedge of spacetime is causally disconnected from the other. It is therefore necessary for him to trace out the field modes which are unobservable. Since we have two such observers which can lie in either regions of Rindler spacetime, two partial traces need to be taken, resulting in one of the four following situations:
\begin{itemize}\item Rob-Rodney, in which the region II modes are traced out for both the $\omega_1$ and $\omega_2$ subspaces. 
\item Rob-AntiRodney, in which region II modes are traced for $\omega_1$ and region I modes are traced for $\omega_2$.
\item AntiRob-Rodney, in which region I modes are traced for $\omega_1$ and region II modes are traced for $\omega_2$.
\item AntiRob-AntiRodney, in which region I modes are traced for both $\omega_1$ and $\omega_2$.
\end{itemize}
In the next section we will study the behaviour of entanglement in all these situations, for a general mixed state of the two field modes under consideration. We note that, unlike in previous works \cite{Edu9,Edu10} where field entanglement between an inertial and an accelerated observer was considered, the inertial entanglement is not  always recovered in the zero acceleration limit of our setting. Although a naive expectation might be that this should be the case, it is not because of the discontinuity in localisation of the Rindler and Minkowski modes: whereas the  former are always localised in region I or II of spacetime, the latter are supported in the whole spacetime.
\section{Entanglement shared by two accelerated partners}\label{res}
We consider a general state described by a $4\times4$ density matrix $\rho$ written in the basis
\begin{align}\mathcal{B}=&\{\ket{0}_{\omega_1,\text{U}}\ket{0}_{\omega_2,\text{U}},\ket{0}_{\omega_1,\text{U}}\ket{1}_{\omega_2,\text{U}},\nonumber\\&\ \ket{1}_{\omega_1,\text{U}}\ket{0}_{\omega_2,\text{U}},\ket{1}_{\omega_1,\text{U}}\ket{1}_{\omega_2,\text{U}}\}.\end{align}
This basis of Unruh excitations is to be expressed in terms of the Rindler basis by means of the results of section \ref{set}. Subsequently, the appropriate partial traces are taken, leaving us with a reduced state $\rho'$ containing all the observable correlations between field modes.

As a distillable entanglement measure valid for mixed states, we employ the negativity \cite{Negat}. From now on when we refer to the negativity of a bipartition we mean the negativity of the corresponding $\rho'$ matrix. 

We will first study the phenomenon of entanglement extinction with a Monte Carlo method to probe the space of density matrices and find out the abundance of states whose entanglement dissapears for finite accelerations.

Then, motivated by the observation that entanglement extinction in the fermionic case appears only for non-pure states,  we will obtain a typical distance (in the sense of degree of purity) between maximally entangled states and the first mixed states where the phenomenon appears.

\subsection{Monte Carlo survey}\label{MCSurv}

The relevant parameter space (4x4 state matrices) is a 15-dimensional manifold, $\mathcal{M}$. We relied on a Monte Carlo method to numerically estimate the abundance of states showing some inertial entanglement which vanishes at infinite acceleration of both partners in the Rob-Rodney bipartition. Our algorithm works as follows: First, we randomly generate a 4x4 matrix  $G$ drawn from the Ginibre ensemble \cite{Ginibre} (normally distributed with identical variance in both real and imaginary parts of all its entries). Then, a state matrix $\rho$ is obtained as 
\begin{align}\rho=\frac{GG^\dagger}{\text{Trace}(GG^\dagger)},\label{ensemble}\end{align} 
and we checked wether if it had some inertial entanglement by  directly computing the negativity of the matrix $\rho$. If so, entanglement at infinite acceleration of both partners was studied.

The use of a matrix $G$ in the Ginibre ensemble, together with the correspondence \eq{ensemble}, induces the Hilbert-Schmidt measure in $\mathcal{M}$  \cite{HSM}, a measure both unitarily invariant (meaning that any two measurable subsets $A_1$ and $A_2$ related through a unitary $U$ such that $A_2=UA_1U^\dagger$ have the same measure) and obtained from a metric. Once we have a measure on $\mathcal{M}$, such questions as `How many states show entanglement survival at infinite acceleration?' become meaningful. Though this unitarily invariant measure is not an unique choice \cite{HSM,otr}, its behaviour is well understood under certain circumstances \cite{otr} and is simple enough to generate states with. 

\begin{figure}[hbtp] 
\includegraphics[width=.50\textwidth]{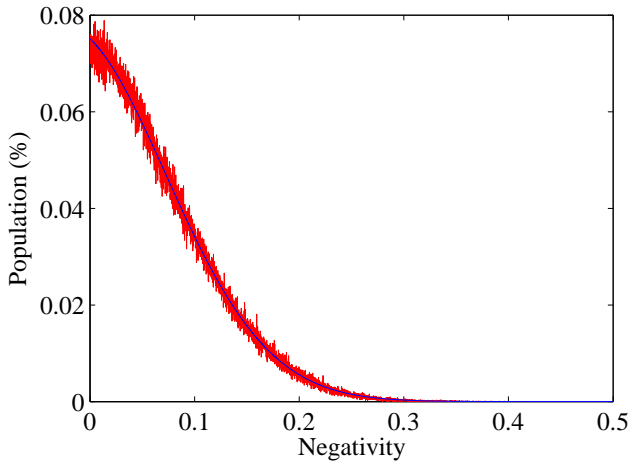}
\caption{(Color online) Negativity histogram for a sample of $10^6$ mixed states drawed from the Ginibre ensemble and corresponding gaussian fit. Note that only entangled states are shown; the fraction of separable states (zero negativity) is much higher (approximately $24\ \%$).} 
\label{histo}
\end{figure}

To gain some physical insight on the behaviour of the induced measure in the space of density matrices, we generated $10^6$ states, plotting an histogram of their inertial negativities, which can be seen in figure \ref{histo}. Approximately $76\ \%$ of these states were entangled, distributing themselves over a gaussian.

We found that from a sample of $10^5$ points, $75.9\ \%$ showed non-zero negativity, and that
\begin{itemize}
\item For $\qr=1$ (the single mode approximation),  $85\ \%$ of the states with nonvanishing inertial entanglement had zero entanglement in the limit of infinite acceleration for both observers.
\item For $\qr=0.97$, $79\ \%$ of the states  with inertial nonvanishing  entanglement had zero entanglement in the limit of infinite acceleration for both observers.
\item For $\qr=1/\sqrt{2}$, $75\ \%$ of the states with nonvanishing  inertial entanglement had zero entanglement in the limit of infinite acceleration for both observers.
\end{itemize}

The results above show that the phenomenon of entanglement extinction at infinite acceleration (and, as we shall see in subsection \ref{graf}, at finite acceleration) is fairly common and not a peculiarity of some specific family of states. Also, the percentage of states showing the phenomenon drops with $\qr$, which is consistent with previous results that going beyond the single mode approximation could result in entanglement amplification \cite{Mig1}.

\subsection{Some states in detail}\label{graf}

\begin{figure}[hbtp] 
\includegraphics[width=.50\textwidth]{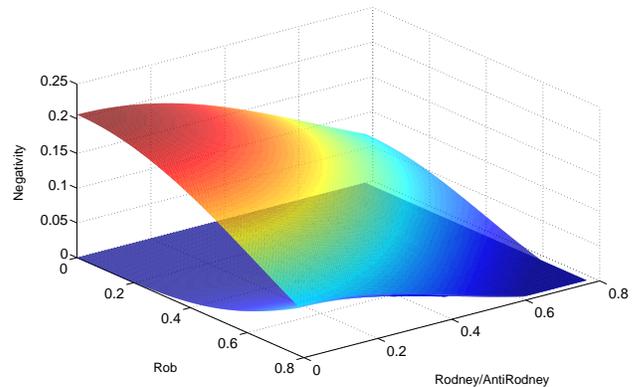}
\caption{(Color online) Negativities of two bipartitions as a function of the acceleration parameters $r_{\omega_1}$ and $r_{\omega_2}$ for the state \eq{ira}. The upper part of the graph corresponds to the Rob-Rodney bipartition, and the lower one, to Rob-AntiRodney. The negativity of the AntiRob-Rodney bipartition is diagonal-symmetric to that of Rob-AntiRodney, and the AntiRob-AntiRodney bipartition is always zero.}
\label{stats}
\end{figure}

In this subsection, some of the states showing the phenomenon of vanishing entanglement at infinite acceleration are analysed thoroughly.  More specifically, we computed the negativity for all four possible bipartitions for the state \eq{ira} which first showed the phenomenon. Considering previous results in \cite{Edu9}, it might have been reasonable to guess that entanglement is not destroyed, but rather transferred to the other bipartitions. This would have been expectable if we considered the results obtained when analysing maximally entangled states. We present our results in figure \ref{stats}.

We remark that, whenever there is entanglement vanishing at some acceleration in the Rob-Rodney bipartition, the entanglement vanishes in all other bipartitions and for all values of acceleration as well. We have thus established that entanglement is not merely being redistributed among the various possible bipartitions, but rather that it is lost. Although true for pure entangled states, it is not true in general that fermionic statistics directly implies entanglement survival. We have seen that for some non-pure fermionic states, entanglement extinction due to acceleration might happen. However, it is true that entanglement survival in the infinite acceleration limit is only possible if there is some entanglement transfer to other bipartitions, as it was concluded in \cite{Edu10}. Note that the correlations between Rob and AntiRodney only show up when $r_{\omega_1}$ is far from $r_{\omega_2}$. In other words, there are only correlations in this bipartition if the accelerations or the  field modes they watch are very different.

The region where entanglement is zero for all bipartitions  does not include points with $r_{\omega_1}=0$ or $r_{\omega_2}=0$. However, for other states, such as the one given by the density matrix
\begin{align}\rho&=\alpha+i\beta, \nonumber\\
\alpha&=\left[ \begin{array}{cccc}
  0.0564 & 0.0190 & 0.0515 & 0.1014 \\
  0.0190 & 0.1902 & 0.0394 & 0.1575 \\
  0.0515 & 0.0394 & 0.2174 & 0.2247 \\
  0.1014 & 0.1575 & 0.2247 & 0.5360
\end{array} \right],\nonumber\\\beta&=
\left[ \begin{array}{cccc}
  0 & -0.0089 & -0.0645 & -0.0339 \\
  0.0089 & 0 & -0.0233 & 0.0149 \\
  0.0645 & 0.0233 & 0 & 0.0962 \\
  0.0339 & -0.0149 & -0.0962 & 0
\end{array} \right].\end{align}
The region of vanishing entanglement does indeed include such points. Since in the single mode approximation the limit of the Unruh modes for zero acceleration yields modes confined to region I, this means that this behaviour is also present where there was only one accelerated observer. The only simple requirement is thus to consider mixed states, as pure states do not present this behaviour.

As mentioned above, there is some entanglement in the Rob-AntiRodney bipartition whenever Rob's acceleration is higher than AntiRodney's and these accelerations are not close. The corresponding result but with AntiRob-Rodney correlations is also true.

We also point out that all the states studied under the single mode approximation show the same qualitative behaviour: Entanglement decreases monotonically along the diagonal up to a threshold value of $r$ where it hits zero, staying there onwards. Higher values of inertial entanglement correspond to higher values of this threshold. It seems likely that the threshold can achieve all the values of $r$ if we allow arbitrarily low inertial entanglement, although we have not proved this.

\begin{figure}[hbtp] 
\includegraphics[width=.50\textwidth]{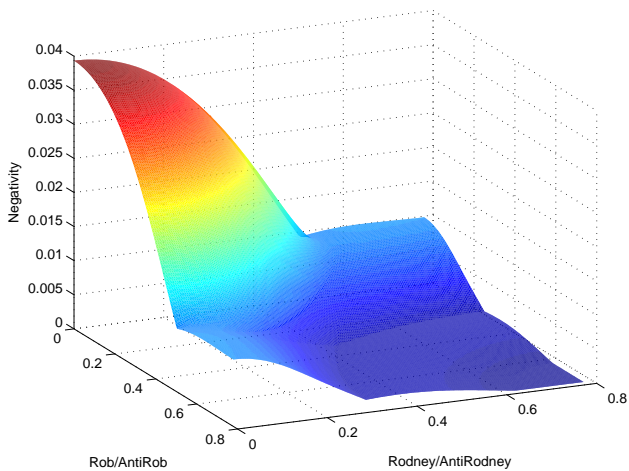}
\caption{(Color online) Negativity of all the bipartitions as a function of the acceleration parameters $r_{\omega_1}$ and $r_{\omega_2}$ for the state \eq{ira} and $\qr=1/\sqrt{2}$. The negativities of all bipartitions are equal since the use of Unruh modes with $\qr=1/\sqrt{2}$ makes regions I and II completely equivalent.}
\label{stats2}
\end{figure}

These results are derived assuming the single mode approximation $\qr=1$. This assumption introduces a strong asymmetry between region I and region II modes, which explains why there are no states showing the same behavior for the AntiRob-AntiRodney bipartition. Figure \ref{stats2} shows  the same plot for negativity as a function of acceleration for state \eq{ira}, when $\qr=1/\sqrt{2}$. The behaviour is qualitatively different once the single mode approximation is relaxed; there can be entanglement amplification maxima (see \cite{Mig1}) and the negativity no longer gives rise to a smooth surface. Also, due to the symmetry between regions present when  $\qr=1/\sqrt{2}$, all four bipartitions result in the same surface.

\subsection{Radius of entanglement survival}
This phenomenon of entanglement destruction at infinite acceleration has only been observed for mixed states. Since maximally entangled states are pure, this means that entanglement survival at infinite acceleration should be expected as long as we do not stray `too far' from maximally entangled states. To determine the precise meaning of `too far', we developed the following scheme: After generating a random state matrix $A$ in the same way as for the Monte Carlo algorithm above, we computed the negativity at $r_{\omega_1}=r_{\omega_2}=\pi/4$ under the single mode approximation as a function of $p$ for the family of states
\begin{align}\rho(p)&=(1-p)\proj{\Psi}{\Psi}+pA,\quad p\in [ 0,1],\label{fam}\\\ket{\Psi}&=\frac{1}{\sqrt{2}}\left(\ket{00}+\ket{11}\right) .\end{align}
until a value of $p$ for which the negativity dropped to zero was found. This defines a function $f:\mathcal{M}\rightarrow \mathbb{R}$ from the space of state matrices to the real numbers. Since $\mathcal{M}$ is compact it follows that $f$ has a minimum value, which may be regarded as a measure of how far from the maximally entangled state of our choice is the set of states with zero entanglement in the infinite acceleration limit.

The process was repeated for $10^6$ random matrices, using the same generating method as in subsection \ref{MCSurv}. The minimum value of $p$ was found to be $p=0.29$.  This means that for any one-parameter family of states of the form \eq{fam}, with a high degree of certainty, the states never become mixed enough to wipe out entanglement at infinite acceleration before this value of $p$ is attained. This further confirms that the phenomenon is linked to non-coherent states, since it does not show up near entangled states.

\section{Conclusions}\label{cojoniak}

We have studied exhaustively the phenomenon reported in \cite{Chinooos,iranies} where it was shown that when one or two observers are accelerating some states show no entanglement survival in the infinite acceleration limit, contrary to previous insights that entanglement survival was the hallmark and a direct consequence of fermionic statistics.

We have found that the phenomenon does not require two accelerated observers to show up,in accordance with \cite{Chinooos}, and that the only fundamental requisite is to consider mixed states. This general study is the first work in relativistic quantum information which explicitly considers aribitrary mixed states from the inertial perspective. We also performed our study beyond the single mode approximation. 

We characterised the phenomenon in a number of ways, finding it to be fairly common in the space of states of 4x4 density matrices, analysing its general features and providing a reasonable measure on the degree of nonpurity necessary to observe the effect. We found that entanglement is not transferred from one bipartition to another, but rather is completely erased.

There are fundamental differences in the way bosonic and fermionic fields behave with acceleration, which have traditionally regarded as having a profound impact on the more detailed aspects of field entanglement. Here we have studied all the possible bipartitions to which entanglement could have relocated,  showing that entanglement completely disappears from our states and thus the phenomenon of fermionic entanglement survival is dependent on the detailed features of the field state, and not only on the form of the change of basis between Minkowski and Rindler bases.

\section{Acknowledgements}
E. M-M was supported by the CSIC JAE-PREDOC2007 Grant scheme. E. M-M and J. L are partially supported  by the Spanish MICINN Project FIS2008-05705/FIS and the QUITEMAD consortium.

%

\end{document}